# Ab-initio Molecular Dynamics study of 1-D Superionic Conduction and Phase Transition in β- Eucryptite

Baltej Singh[1,2], Mayanak K. Gupta[1], Ranjan Mittal[1,2] and Samrath L. Chaplot[1,2]

[1]Solid State Physics Division, Bhabha Atomic Research Centre, Mumbai, 400085, India
[2]Homi Bhabha National Institute, Anushaktinagar, Mumbai 400094, India

## Abstract

β- Eucryptite ($LiAlSiO_4$) is a potential electrolyte for Li- ion battery due to its high Li- ion conductivity and very small volume thermal expansion coefficient. We have performed ab-initio molecular dynamics simulations of β- Eucryptite to study the origin of high temperature superionic phase transition in this material. The simulations are able to provide the microscopic understanding of the one -dimensional superionicity that occurs along the hexagonal c-axis and is associated with the order-disorder nature of the phase transition. The Li ionic conductivity is found to enhance due to the anisotropic negative thermal expansion along hexagonal c-axis. The introduction of defects in the crystal like, excess Li in interstitial sites, Li vacancy and O vacancy are found to significantly increase the ionic conductivity and hence might reduce the temperature of the superionic phase transition in this material.

# I. INTRODUCTION

Clean and green storage of energy is achievable by various energy storage technologies[1-4]. Batteries are one of the most eligible devices[5-8] for this purpose. The solid state batteries[6, 9-12] offer a most compact and safe version to store energy at various scales. These batteries contain a solid electrolyte between two electrode materials and hence no separator is required to fabricate these batteries. The solid electrolyte[13], being more stable[6, 9-11] than liquids, could resolve the issue of leakage, chemical stability, vaporization, flammability and dendrites formation to a great extent. It will also enable the use of metallic Li as anode material[9, 14] and will enhance the current charge capacities.

However, solid electrolytes possess a higher activation energy[6, 10] for ionic diffusion through their body and hence offer smaller ionic conductivity values as compared to that of liquid and salt-polymer based electrolytes[15]. Inflation in the magnitude of Li ionic conductivity in solid electrolytes[6-8, 10] at room temperature is essential for engineering highly efficient[16] solid state batteries for hybrid vehicles applications. An enormous amount of research is underway to investigate and engineer a best suitable solid electrolyte for this purpose. Although efforts are being made to design solid ion conductors containing various small alkali[2, 6, 7, 10, 11, 17] ($Li^+$, $Na^+$ etc.) and alkaline earth ions ($Mg^{+2}$), yet only Li- based structures dominated in quantity and quality due to the smallest ionic radius and highest ionic mobility of Li. The recently discovered $Li^+$- based solid electrolytes have an ionic conductivity of the order of 0.0001 to 0.1 $Ohm^{-1}$ $cm^{-1}$ and activation energy of about 0.1- 0.5 eV, and are known as superionic Li conductors. Sulphides and oxides are the main classes[11, 18-21] of materials possessing a liquid- like Li ionic conduction. Oxide electrolytes[13] possess slightly lower ionic conductivity as compared to sulphides, but they can be used in high voltage batteries due to their wide range of stability in comparison to that of sulphides.

The experimental techniques like ionic conductivity measurements and NMR spectroscopy[22, 23] are useful tools to characterize the solid ionic conductors and provide the activation barriers for Li diffusion. The pathways for atomic diffusion can be obtained from the temperature dependent X-ray[24] and neutron diffraction techniques[25]. However, Li atom, due to its small atomic number (Z=3), is known to be a poor scatterer of X-ray. Even for neutron, Li has low scattering cross section (= 1.37 barns) and high absorption cross section (= 70.5 barns), which makes it difficult to study the behaviour of Li using neutron scattering. Computational tools, like density functional theory and molecular dynamics simulations, provide a very good alternative to study the atomic scale behaviour of Li atom[18, 19]. The fundamental unsolved problem is to find the atomistic origin of high ionic conductivity in lithium superionic conductors and solving this will help to design new and improved solid electrolyte materials. Molecular dynamical simulation is a wonderful tool to study the dynamical properties[26-28] like diffusion constant, the activation energy, and the diffusion path for solid electrolytes at various temperature/pressure conditions. The empirical potential based molecular dynamical simulations are computationally cheap but sometimes they do not give the actual behaviour[29] of the system unless accurate potential parameters are considered. The advancement in computer technology in recent years has enabled to perform the ab-initio molecular dynamical simulations for hundreds of atoms. This technique is widely being used to study and design a number of suitable solid electrolytes for all solid-state batteries.

Topology and geometry[30-32] play an important role in deciding the ionic conductivity of Li in solid electrolytes. Based on this, solid electrolytes can have isotropic (3-D), planner (2-D) and channel-like (1-D) Li diffusion pathways. One- dimensional solid electrolyte provides highly directional effective diffusion paths perpendicular to the surface area of electrode materials, hence avoids the unnecessary diffusion along the surface. However, only a few one- dimensional solid electrolytes[33, 34] like $Li_{10}GeP_2S_{12}$, $Li_3M(MoO_4)_3$ (M = Cr,Fe), $MgVPO_4F$ have been known to possess superionic diffusion.

$LiAlSiO_4$ (β- eucryptite) is a very important one- dimensional superionic conductor. It is one of the most promising and unique candidates which possess superionic conductivity along with anisotropic negative thermal expansion (NTE) behaviour[20, 29, 35]. $LiAlSiO_4$ has a hexagonal structure[36] (space group $P6_222$) that expands in the ab-plane and contracts along the c- axis upon heating. It shows an overall very low volume thermal expansion[37, 38] and has good mechanical stability. These properties make it a potential electrolyte for Li ion batteries[39]. Lithium is decorated in one-dimensional channels parallel to the c-axis made up from the double helices of $SiO_4$ and $AlO_4$ tetrahedra arrangement. There are alternating double layers of Al and Si normal to the hexagonal c-axis. Each channel has six available tetrahedral sites; however, at ambient temperature, only three of the available sites are occupied[40] in an alternating sequence. The high- temperature phase transition occurs by disordering of Li among all the available sites along c-axis. The high temperature phase[36] (space group $P6_222$) exhibits a one dimensional superionic Li conduction. Earlier classical molecular dynamical simulations[29] for $LiAlSiO_4$ could not quantitatively reproduce the experimentally observed superionic temperature and anisotropy in Li ionic conduction.

We have now carried out extensive ab-inito molecular dynamical (AIMD) simulations to study the mechanism of the high- temperature (HT) phase transition and distribution of Li in the high temperature structure. We also studied the mechanism of superionic conduction in the HT phase and energy barrier for Li diffusion. The effect of negative thermal expansion (NTE) on superionic conduction is interpreted for the first time. Moreover, we have studied various types of defects which give rise to inflation in superionic conductivity and might reduce the temperature for the phase transition.

**II. Computational Details:**

We have performed ab-initio molecular dynamics simulations (AIMD) as a function of temperature for $LiAlSiO_4$. The Vienna based ab-initio simulation package (VASP) was used for the calculations[41, 42] of the structure and dynamics. The calculations are performed using the projected augmented wave (PAW) formalism of the Kohn-Sham density functional theory within generalized gradient approximation (GGA) for exchange correlation following the parameterization by Perdew, Becke and Ernzerhof.[43, 44]. The plane wave pseudo-potential with a plane wave kinetic energy cut off of 820 eV was adopted. The integration over the Brillouin zone is sampled using a k-point grid of size 2×2×2, generated automatically using the Monkhorst-Pack method[45]. The above parameters were found to be sufficient to obtain a total energy convergence of less than 0.1 meV for the fully relaxed (lattice constants & atomic positions) geometries. The total energy is minimized with respect to structural parameters.

The thermal expansion of $LiAlSiO_4$ is obtained from Parrinello- Rahman (NPT) dynamics[46, 47] with Langevin thermostat[48]. The Newton's equations of motion are integrated using the Verlet's algorithm with a time step of 2 fs. A unit cell of 84 atoms for the room temperature (RT) phase and a supercell of 189 atoms, equivalent to 3×3×1 supercell of the high-temperature phase (21 atoms per unit cell), subject to the periodic boundary conditions are used. Ab-initio molecular dynamics simulations are computationally expensive, so we have taken a supercell of 3×3×1 for our calculations. Another reason for choosing 3×3×1 supercell for simulations is that this cell contains large number of Li channels along the c-axis which allows to study the interchannel correlations. That is, how diffusion of Li in a channel is affected by diffusion of Li in other channels. Additional simulations are performed in a 2×2×2 supercell of the high-temperature phase at 1000 K.

Further, we have studied RT to HT phase transition in $LiAlSiO_4$ and lithium diffusion in the high temperature (HT) phase. The temperature in NVT simulations is attained through Nosé thermostat[49]. Initially, the structure was equilibrated for 2 ps to attain the required temperature in NVT simulations. Then the production runs upto 10 ps in NVE simulation are performed (the temperature remained essentially constant). Simulations are performed for a series of temperatures from 300 to 1200 K. At each temperature, a well-equilibrated configuration is observed during the 10 ps simulation.

## III. Results and Discussion

### A. Phonon Spectra & Thermal Expansion

Phonon density of states of a compound gives the spectrum of vibrational modes of atoms in the solid and hence provides information about the strength and nature of bonds in the solid. Although quasiharmonic lattice dynamics is a good tool[50] to calculate the phonon spectrum, it does not include the anharmonic contribution coming from the temperature. The ab- initio molecular dynamics directly encounters the anharmonic effect[51, 52] and hence provides accurate determination of phonon spectrum at any temperature. We have calculated and compared the phonon spectrum (Fig. 1) of β- eucryptite at ambient temperature (for RT phase) and pressure with the available inelastic neutron scattering measurements[35]. The phonon spectrum is calculated from the Fourier transformation of the velocity–velocity autocorrelation function as given below:

$$g(\omega) = \int e^{i\omega t} \frac{\langle \mathbf{v}(t)\mathbf{v}(0)\rangle}{\langle \mathbf{v}(0)\mathbf{v}(0)\rangle} dt \tag{1}$$

Where the angle brackets <> indicating an ensemble average over various atoms. The velocity autocorrelation function is obtained from the trajectories of all the atoms recorded during the molecular dynamics simulations. The peak structure in the calculated phonon spectrum (Fig. 1) agrees very well with the experimental spectrum. Moreover, both the spectra extend up to the same maximum energy. A very small underestimation in the calculated stretching modes comes from the slightly overestimated bond lengths in the framework of GGA approximation. The overall match of the phonon spectra validates our simulations cell and other input parameters for further calculations of various thermo dynamical quantities.

Our previous work[35] on this compound was based on lattice dynamics calculations of thermal expansion behaviour under quasiharmonic approximation. The method works very well at low temperature, however at high temperature the calculated linear thermal expansion is underestimated. This underestimation at high temperature could be result of anharmonic vibration of atoms, which could not be taken care in the quasiharmonic approximation. The ab-initio molecular dynamics method is known to include the anharmonic effect at high temperature. The calculations upto 800K reproduce the experimental results (Fig.2) very well, which show contraction along the c-axis and expansion along the a- axis.

B. **High temperature Phase transition in *β*-eucryptite**

The room temperature phase of *β*-eucryptite[40] (Fig. 2) contains 12 *Li* atoms which are present at three different Wyckoff sites namely *3b, 3c* and *6f*. These *Li* atoms are distributed in the channels along the hexagonal c-axis. There are two types of channels[40], namely, '*S*' type and '*A*' type. In '*S*' type channel *Li* atoms are coplanar with the *Si* atoms while in '*A*' type channel *Li* atoms are coplanar with the *Al* atoms. There are three secondary (*S*) and one primary (*A*) channel in a single unit cell of the room temperature structure[40]. Another difference between these channels is that the '*A*' type channel contains Li which are present at *3b(x,y,z)* Wyckoff site and is named as *Li1*, while the '*S*' type channels have a distribution of *Li* over 3c(x,y,z) and 6f(x,y,z) sites and are named as *Li2* and *Li3* respectively. Moreover when '*A*' type channel is moved by *c/6* along the hexagonal c-axis, it becomes '*S*' type channel[40].

The mean square displacements of various atoms (Li, O, Al & Si) are calculated from the time evolution of the trajectory of these atoms at room temperature (300K). These calculations are performed in micro canonical ensemble (NVE) and the volume of unit cell is held constant to that of relaxed structure at 0K. The temperature, pressure and energies are well converged in these calculations. The calculated mean square amplitudes (Fig. 3a) show that the Li being lightest element has the highest but Al and Si being heavier has lower mean square amplitudes. Fig 3a show that all the atoms are vibrating about their equilibrium positions with small amplitudes of vibrations and hence the structure is stable in this configuration.

As simulation temperature is increased to 600K, we see (Fig 3b & Fig 4b) that although Al, Si, and O atoms have very small and almost constant values of mean square displacements, Li atom show a sudden increase after 4 ps. We analysed the anisotropic mean square displacements (Fig 3c & Fig 4b) for Li atoms and found that the MSD's show an anomalous value only along the hexagonal c-axis while it is negligible in the *ab-* plane. This implies that Li atoms start move along the hexagonal c-axis with a jump like behaviour. When all individual atom's trajectory (Fig 3d) are analysed, we found that three Li atoms are moving altogether. These groups of three Li atoms belong to the same channel. The very first channel movement (Fig 3d) occur around a simulation time of 4 ps. This is found to be the 'A' type channel. All the Li atoms in this 'A' type channel are found to have mean square displacement of ~3.6 Å$^2$ which corresponds to a displacement of ~1.9 Å. This displacement is equal to c/6. This indicates that the 'A' type channel is transformed to the 'S' type channel. Afterwards, the Li atoms (three atoms together) in other channels also start moving along c-axis with displacement amplitude of about c/6 and c/3 etc. Hence Li atoms have a possibility to be present in all the available sites along the channels at 600K. All the channels become equivalent with a distribution of Li atoms at all the available sites. This gives rise to the high- temperature phase transition in *β*-eucryptite[36]. This transition was experimentally reported[36, 53] at around 698-758 K. In the high temperature phase, all the channels become equivalent and the hence the new symmetry reduces the 'a' (or 'b') lattice parameter to half of their original values. Experimentally[53] the high temperature structure of *β*-eucryptite is also known to have Li atoms positional disordering along the hexagonal c-axis.

The favourable movement of Li atoms in the 'A' type channel in comparison to that in the 'S' type channel may be statistical or may be related to their activation energy barriers. To analyse the distinction between the 'A' and 'S' type channels, we have calculated the activation energy barrier for motion of Li atoms in these channels using climbing image nudged elastic band method. We found that the activation energy barrier for both types of channels is almost equal. This indicates that activation energy provides an equal probability for movement of both types of Li channels. However, the calculated energy shows that the final state obtained after moving channel A is energetically more favourable in comparison to that for S type channel by an enthalpy difference of 0.35 meV.

Further, we have calculated the total enthalpy of final structural configurations (TABLE I) obtained from the shift of A and S type channels by c/6 along the c-axis. $N_A$ and $N_S$ represent the number of A and S type Li channels respectively in single unit cell containing 84 atoms. The calculated enthalpy is compared for the following configurations: (i) when $N_S$: $N_A$:: 3: 1, this corresponds to actual room temperature phase unit cell, (ii) when all Li atoms are coplanar with Si atoms ($N_S$: $N_A$:: 4: 0), this is obtained by shifting all Li atoms in A- type channel by c/6 along c-axis, (iii) when $N_S$: $N_A$:: 2: 2, there are equal number of Li atoms in Al and Si planes, (iv) when $N_S$: $N_A$:: 1: 3, and (v) when $N_S$: $N_A$:: 0: 4, (vi) Li occupying three consecutive sites at 0, c/6 and c/3 in one of the four channels while all other Li are present at their equilibrium positions, (vii) Li occupying three consecutive sites at 0, c/6 and c/3 in all of the four channels. The calculated (0 K) enthalpy and lattice parameters for various possible combinations of Li occupancy in the channels are given in (Table I). The RT Phase possesses the minimum enthalpy and hence represents the most stable configuration. Moreover, some of the other possible configurations have fairly low enthalpy. Hence Li in the channels of the high temperature phase may occur at all the available sites with some finite probability. The high temperature phase is characterised by the statistical distribution of Li atoms in all the available sites in the hexagonal channels. The Li atoms seem to be disordered in channels and the high temperature phase transition may be of order- disorder nature[53].

If an ordered structure is assumed for the HT phase (configuration (ii) TABLE I) it gives rise to a calculated c parameter which is larger than that of the RT phase. However, when various possible configuration ((iii), (iv) & (v) TABLE-I) of structure with Li at different sites are considered we get lower *c* lattice parameter. Hence, Li in the channels of the high temperature phase would have a distribution over all the sites. This implies that the HT-phase transition occurs by disordering of Li atoms and produces little change in the lattice parameters. We have also calculated the phonon spectra of β- Eucryptite for structural parameters close to the Li-diffusion temperature (800 K). However we didn't get any unstable phonon mode implying that phonons do not contribute to this transition.

**C. Superionic Conductivity in HT-Phase of *β*-eucryptite**

The simulations performed at 800 K indicate increase in Li channel movements. At 800K (Fig 5a & Fig 6a), the compound is already in the high temperature structure, so all the Li containing channels are equivalent. The channel movement begins at smaller simulation time in comparison to that for 600K, and some channels start moving coherently at same time. The increasing mean squared displacements (MSD's) of Li atoms in these channels signify the diffusion in Li channels along the hexagonal c-axis. When temperature is further raised to 1000K (Fig 5b & Fig 6b), more numbers of channels are seen to be diffusing at even smaller simulation times. During the diffusion, there is correlation between the Li atoms in each channel which can be seen at 1000K (Fig 5b & Fig 6b). The diffusion of Li channels in this compound is found to be one- dimensional jump- like rather than continuous diffusion. At sufficiently high temperature of 1200 K (Fig 6c), some of the Li atoms also started diffusing in the ab-plane; hence at high temperatures the intra- channel correlation of Li atoms decreases.

For the calculation of diffusion coefficient, we have averaged the mean-square displacements (MSD) over all the Li atoms and over a large number of time origins. This averaging gives rise to disappearance of step like behaviour in the MSD curve. The slope of the MSD versus time plot is used to get the diffusion coefficients in accordance with Einstein's formula

$$2Dt = \frac{1}{N_D} \langle |r_i(t) - r_i(0)|^2 \rangle \qquad (2)$$

where $r_i$ is the position vector of the diffusing species, t is the time and $N_D$ represents the dimensionality of diffusion. $N_D$ =1 in our case since diffusion is occurring only along the c-axis. The quantity within the brackets has been averaged over different initial times.

We have calculated the average diffusion coefficient for all the Li atoms in the supercell. The calculated diffusion coefficient (Fig 7a) shows a sharp increase in magnitude around 600K. The calculated value of diffusion coefficient has magnitude ($\approx 10^{-9}$ m$^2$s$^{-1}$) comparable to that of atoms in liquids. So, the high temperature structure of *β*-eucryptite possesses a superionic Li conduction which is one dimensional in nature. The jump distances and resting time are reflected in the mean square displacements as a function of time (Fig 3 &5). Typical jump distances along hexagonal c-axis are calculated to be 1.9 and 3.8 Å which corresponds to c/6 and c/3 jumps. The resting time decreases with increase in temperature, and typical resting time at 600 K varies from 2 to 4 ps (Fig 3). The jump distances correlate to the structure because there are filled and vacant Li sites along the channel at c/6 separation. The calculated jump distances are consistent with the experimental quasi-elastic neutron scattering results[54].

We have also performed a simulation in a 2×2×2 supercell at 1000 K (Fig. 5(c)), and the results are found to be consistent with that in a 3×3×1 supercell. This may be expected since the c-axis is fairly long (11.48 Å) and each Li-site has a partial occupancy of only 0.5 in the super-ionic phase. The partial occupancy provides the necessary freedom for the Li-motion.

The activation energy for Li jump diffusion can be obtained from the Arrhenius plot of diffusion coefficient as a function of temperature as given by [55]

$$D = D_0 e^{-\frac{E_a}{k_B T}} \quad (3)$$

Where $D_0$ is the pre- exponential factor, $k_B$ is the Boltzmann constant and T is the temperature. The logarithm of this equation is

$$\ln(D) = \ln(D_0) - \frac{E_a}{k_B T} \quad (4)$$

The calculated activation energy (Fig 7b) for diffusion of Li atoms is 0.3 eV which agrees with those calculated from the nudged elastic band method (0.3-0.4 eV) calculations[29] and with the experimental[56] values of 0.6-0.8 eV.

The pair distribution function (PDF) for various atomic pairs of *β*-eucryptite is calculated (Fig 8) as a function of temperature. The PDF for Al-O and Si-O pairs does not seem to change with temperature. This indicates that Al-O and Si-O bonds remain nearly rigid up to high temperatures of 1200 K. On the other hand, the PDF for Li related atomic pairs are highly affected with increasing temperature. This gives a signature that at high temperature the superionic phase transition in *β*-eucryptite mainly arises from the dynamical behaviour of Li atoms.

The polyhedral angles (Fig 9) as a function of temperature are calculated from the time evolution of structure as obtained from molecular dynamical simulations. At 300 K, the O-Al-O bond angles have an average value of $109.2^0$. The O-Si-O bonds exhibit two peaks around $113.8^0$ and $100.6^0$ corresponding to two types of bond angles around Si atoms. The spread in bond angles shows that the $AlO_4$ are $SiO_4$ polyhedral units are irregular in nature. However, as a function of temperature the double peak structure of O-Si-O bonds merges to single peak. With increase in temperature, the configurational entropy increases and the structure has more possible values of O-Si-O angles. This in turn creates a broad distribution of O-Si-O bond angles and the two peak structures disappear in this broadening. The total spread of O-Al/Si-O bond angles does not change significantly with increasing temperature. This implies that the irregularity of $(Al/Si)O_4$ polyhedral units do not increase at high temperature. However, the bond angles Si-O-Al show significant increase in spread with increase in temperature. This indicates increase in flexibility for the rotation of corner shared $AlO_4$ and $SiO_4$ polyhedra about the shared O atom. This would give rise to a flexible network which favours the movement of Li atoms. Li atom bond with O atoms shows two types of widely spread O-Li-O angles showing a highly irregular polyhedral nature. The two O-Li-O bond angles show some spread with temperature, but the two peak structure is preserved. This suggests that the Li- jumps are quite fast and Li remains at definite positions (0,c/6,c/3,c/2,2c/3,5c/6) at most of the time.

**D. Relation of NTE and Li diffusion in HT-Phase of *β*-eucryptite**

Due to the higher computational cost for performing ab- initio NPT simulations, we have used instead micro canonical (NVE) ensemble approach to compute the Li diffusion in *β*-eucryptite. As these studies are performed at fixed lattice parameters corresponding to the relaxed 0 K structure, so the effect of thermal expansion on Li diffusion was not considered. *β*- eucryptite is a unique superionic conductor exhibiting the anisotropic negative thermal expansion (Fig 10 a) behaviour. The negative thermal expansion and superionic conduction in *β*-eucryptite are temperature driven phenomenon. So, there is a need to find the relation between these two phenomena. In order to see the effect on Li diffusion due to the anisotropic negative thermal expansion behaviour we have performed a series of ab-initio MD simulations in NVE ensemble at 800 K for calculated structure with $c_0/a_0$ (as at 0 K), $0.99c_0/a_0$ and $0.98c_0/a_0$. These values of c/a reflect the actual thermal expansion behaviour of *β*-eucryptite as a function of increasing temperature. The study is performed in a well determined (HT-phase) superionic state of *β*-eucryptite.

We have calculated the mean square displacement (Fig 10 b, MSD integrated over different initial times) as a function of simulation time for structure with different c/a ratios. Fig 10(b) shows

MSD averaged over all the Li atoms and implies that MSD value is enhanced by decreasing c/a ratio. We have further analysed the Li channel motions in structures with different c/a ratios. The magnitudes of MSD for Li channels diffusion are found (Fig 11) to increase with decreasing c/a ratio. The increase in diffusion with decreasing c/a ratio happens since expansion in the ab-plane opens the free space for Li movement along the c-axis. Also, the contraction of the c-axis also reduced the available jump distance along the c- axis.

We have also calculated the activation energy for Li channel diffusion in all the considered structure configurations using static climbing image nudged elastic band method (Fig 12). The increase in diffusion with decreasing c/a ratio comes from the decrease in activation energy for Li channel diffusion.

**E. Factors affecting Li Diffusion in HT-Phase of *β*-eucryptite**

There may be many factors affecting ionic motion in a crystal. We have considered three type of defects in the crystal to study Li diffusion. These defects are (i) Li excess (ii) Li vacancy and (iii) O vacancy. This study is also performed at T=800K which corresponds to a well determined superionic state of *β*-eucryptite.

**Li Excess:** For Li excess case, a Li atom is inserted in the hexagonal channel in a vacant tetrahedral site of 3×3×1 supercell of HT phase structure. The system is equilibrated for 3 ps and the excess Li is found to settle at position in between the already present Li atoms along c- axis, but slightly displaced in ab-plane from the channel along the c-axis. The molecular dynamical simulations (NVE) are performed for 15 ps and MSD (averaged over all Li atoms) from these simulations are plotted in Fig 10. Contrary to the earlier cases, Li atoms also show movement in the ab-plane. However, even in this case, the diffusion of Li is dominated along the hexagonal c-axis.

The atom- wise analysis of the MSD for Li atoms along the various directions is shown in Fig 12 a. MSD jumps along the a-axis and another perpendicular direction in the ab-plane show that there are three Li atoms diffusing in the ab-plane and coming back to their original positions. Among these three Li atoms, two Li atoms belong to the channel where excess Li atom was inserted and the third Li atom belongs to its nearest channel. It can be seen that the movement of Li in ab-plane does not give any net diffusion, however it only decouples (Fig 13) the correlated motion of Li atoms in a channel. Due to this decoupling, the Li atoms which were moving altogether with equal MSD earlier, now show different MSD value along c- axis. Therefore the presence of excess Li atom in the tetrahedral void in *β*-eucryptite gives rise to extra diffusion and decouples the correlated dynamics of Li atoms in channels along the hexagonal c- axis.

**Li Vacancy:** A Li vacancy is created at the centre of 3×3×1 supercell of the HT phase of *β*-eucryptite by removing a Li atom. Now there are 26 Li atoms in the simulation cell. The structure is equilibrated to obtain a temperature of 800 K in NVE ensemble. The simulation shows that the Li atoms in the vacancy-containing channel come slightly closer to the central vacancy. The simulation results (Fig 14) for MSD as a function of time show that there is no translation or diffusional motion of Li in the ab-plane. However, diffusion of Li is observed along the hexagonal c-axis like that in a perfect HT-phase. The diffusion is found to be enhanced in magnitude in comparison to that in the perfect crystal. Li atoms in the vacancy- containing channel possess the largest MSD and initialize the diffusion process in the structure. We observe that the intra- channel correlation between the diffusing Li atoms is decreased as compared to that in the perfect crystal. This might happen because of the increased freedom of motion of Li atoms in vacancy-containing channel due to availability of more number of tetrahedral sites when Li vacancy is created.

**O Vacancy:** Several oxides are found to be oxygen deficit[57-59] even in normal synthesis processes although oxygen deficiency can be easily introduced by synthesis in a reduced environment. Here we have calculated the effect of oxygen vacancy on Li diffusion. Oxygen vacancy was created near the centre of 3×3×1 super cell of the HT phase of *β*-eucryptite by removing an O atom from 144 O atoms. The structure is equilibrated in NVE ensemble and then a production run is performed. The calculated anisotropic MSDs of all Li atoms in the structure are plotted (Fig 15) as a function of time. It can be seen that similar to the perfect crystal there is no diffusion of Li in ab-plane. However, the magnitude of diffusion along the hexagonal c-axis increases substantially in comparison to that of the perfect crystal. Moreover, there is still very good intra- channel correlation between the Li atoms in all the channels. Oxygen vacancy enhances the diffusion of Li without interrupting the intra- channel correlation.

**IV. Summary**

We have performed a thorough study of solid electrolyte β-eucryptite (LiAlSiO$_4$) using ab-initio molecular dynamics simulations. The computed phonon density of states and temperature evolution of lattice parameters agree very well with the available experimental data. The computed mean square displacements show a transition to the high- temperature phase at around 600K initiated by the movement of Li atoms in the A type channel. The high- temperature phase is found to have disordered Li atoms over all the six available sites in each channel of the unit cell. The reason for Li disorder in HT phase is elucidated by the calculated enthalpy of various possible structures. The compound is found to exhibit a one-dimensional Li superionic conduction in the HT phase along the hexagonal c-axis. The correct nature of anisotropy in superionicity is very well reproduced by ab- initio calculations as compared to somewhat less satisfactory results obtained from classical molecular dynamical simulations[29]. The anisotropic negative thermal expansion along the hexagonal c-axis is found to enhance Li ionic conductivity along the c-axis. The introduction of defects in the crystal, namely, Li interstitial, Li vacancy and O vacancy, are found to increase the ionic conductivity and hence might reduce the temperature of HT superionic phase transition in this material. The synthesis of the structure with Li or O atoms vacancies might provide a possibility to achieve a nearly room temperature solid electrolyte.

**Acknowledgements**

S. L. Chaplot would like to thank the Department of Atomic Energy, India for the award of Raja Ramanna Fellowship.

1. J.-H. Lee, J. Kim, T. Y. Kim, M. S. Al Hossain, S.-W. Kim and J. H. Kim, *Journal of Materials Chemistry A*, 2016, **4**, 7983-7999.
2. X. Luo, J. Wang, M. Dooner and J. Clarke, *Applied Energy*, 2015, **137**, 511-536.
3. M. R. Lukatskaya, B. Dunn and Y. Gogotsi, *Nature Communications*, 2016, **7**, 12647.
4. H. Sun, Y. Zhang, J. Zhang, X. Sun and H. Peng, *Nature Reviews Materials*, 2017, **2**, 17023.
5. K. C. Divya and J. Ostergaard, *Electric Power Systems Research*, 2009, **79**, 511-520.
6. W. Zhou, Y. Li, S. Xin and J. B. Goodenough, *ACS Central Science*, 2017, **3**, 52-57.
7. J. B. Goodenough and K.-S. Park, *Journal of the American Chemical Society*, 2013, **135**, 1167-1176.
8. D. Larcher and J. M. Tarascon, *Nat Chem*, 2015, **7**, 19-29.
9. M. H. Braga, N. S. Grundish, A. J. Murchison and J. B. Goodenough, *Energy & Environmental Science*, 2017, **10**, 331-336.
10. C. Sun, J. Liu, Y. Gong, D. P. Wilkinson and J. Zhang, *Nano Energy*, 2017, **33**, 363-386.
11. Y. Kato, S. Hori, T. Saito, K. Suzuki, M. Hirayama, A. Mitsui, M. Yonemura, H. Iba and R. Kanno, *Nature Energy*, 2016, **1**, 16030.
12. J. Janek and W. G. Zeier, *Nature Energy* ,, 2016, **1**, 16141.
13. V. Thangadurai, S. Narayanan and D. Pinzaru, *Chemical Society Reviews*, 2014, **43**, 4714-4727.
14. J. Y. Hwang, S. T. Myung and Y. K. Sun, *Chemical Society Reviews*, 2017, **46**, 3529-3614.
15. L. Yue, J. Ma, J. Zhang, J. Zhao, S. Dong, Z. Liu, G. Cui and L. Chen, *Energy Storage Materials*, 2016, **5**, 139-164.
16. E. D. Wachsman and K. T. Lee, *Science*, 2011, **334**, 935.
17. M. Uitz, V. Epp, P. Bottke and M. Wilkening, *Journal of Electroceramics*, 2017, 1-15.
18. G. K. Phani Dathar, J. Balachandran, P. R. C. Kent, A. J. Rondinone and P. Ganesh, *Journal of Materials Chemistry A*, 2017, **5**, 1153-1159.
19. J. Yang and J. S. Tse, *Computational Materials Science*, 2015, **107**, 134-138.
20. U. V. Alpen, E. Schönherr, H. Schulz and G. H. Talat, *Electrochimica Acta*, 1977, **22**, 805-807.
21. N. Kamaya, K. Homma, Y. Yamakawa, M. Hirayama, R. Kanno, M. Yonemura, T. Kamiyama, Y. Kato, S. Hama, K. Kawamoto and A. Mitsui, *Nat Mater*, 2011, **10**, 682-686.
22. O. Pecher, J. Carretero-González, K. J. Griffith and C. P. Grey, *Chem. Mater*, 2017, **29**, 213-242.
23. C. P. Grey and N. Dupré, *Chemical Reviews*, 2004, **104**, 4493-4512.
24. J. Bréger, N. Dupré, P. J. Chupas, P. L. Lee, T. Proffen, J. B. Parise and C. P. Grey, *Journal of the American Chemical Society*, 2005, **127**, 7529-7537.
25. D. Wiedemann, S. Indris, M. Meven, B. Pedersen, H. Boysen, R. Uecker, P. Heitjans and M. Lerch, *Zeitschrift für Kristallographie-Crystalline Materials*, 2016, **231**, 189-193.
26. Z. Zhu, I.-H. Chu and S. P. Ong, *Chemistry of Materials*, 2017, **29**, 2474-2484.
27. C. Chen, Z. Lu and F. Ciucci, *Scientific reports*, 2017, **7**, 40769.
28. X. Wang, R. Xiao, H. Li and L. Chen, *Physical Review Letters*, 2017, **118**, 195901.
29. B. Singh, M. K. Gupta, R. Mittal, M. Zbiri, S. Rols, S. J. Patwe, S. N. Achary, H. Schober, A. K. Tyagi and S. L. Chaplot, *Physical Chemistry Chemical Physics*, 2017, **19**, 15512-15520.
30. E. Kendrick, J. Kendrick, K. S. Knight, M. S. Islam and P. R. Slater, *Nat Mater*, 2007, **6**, 871-875.
31. P. Goel, M. K. Gupta, R. Mittal, S. Rols, S. J. Patwe, S. N. Achary, A. K. Tyagi and S. L. Chaplot, *Journal of Materials Chemistry A*, 2014, **2**, 14729-14738.
32. M. K. Gupta, P. Goel, R. Mittal, N. Choudhury and S. L. Chaplot, *Physical Review B*, 2012, **85**, 184304.
33. Y. Mo, S. P. Ong and G. Ceder, *Chemistry of Materials*, 2011, **24**, 15-17.
34. J. Wu, G. Gao, G. Wu, B. Liu, H. Yang, X. Zhou and J. Wang, *RSC Advances*, 2014, **4**, 15014-15017.

35. B. Singh, M. K. Gupta, R. Mittal, M. Zbiri, S. Rols, S. J. Patwe, S. N. Achary, H. Schober, A. K. Tyagi and S. L. Chaplot, *Journal of Applied Physics*, 2017, **121**, 085106.
36. H. Guth and G. Heger, *North-Holland,* , 1979.
37. L. Xia, G. W. Wen, C. L. Qin, X. Y. Wang and L. Song, *Materials & Design*, 2011, **32**, 2526-2531.
38. H. Xu, P. J. Heaney, D. M. Yates, R. B. Von Dreele and M. A. Bourke, *Journal of Materials Research*, 1999, **14**, 3138-3151.
39. R. T. Johnson, B. Morosin, M. L. Knotek and R. M. Biefeld, *Physics Letters A*, 1975, **54**, 403-404.
40. W. W. Pillars and D. R. Peacor, *American Mineralogist*, 1973, **58**, 681-690.
41. G. Kresse and J. Furthmüller, *Computational Materials Science*, 1996, **6**, 15-50.
42. G. Kresse and D. Joubert, *Physical Review B*, 1999, **59**, 1758-1775.
43. J. P. Perdew, K. Burke and M. Ernzerhof, *Physical Review Letters*, 1996, **77**, 3865-3868.
44. J. P. Perdew, K. Burke and M. Ernzerhof, *Physical Review Letters*, 1997, **78**, 1396-1396.
45. H. J. Monkhorst and J. D. Pack, *Physical Review B*, 1976, **13**, 5188-5192.
46. M. Parrinello and A. Rahman, *Journal of Applied physics*, 1981, **52**, 7182-7190.
47. M. Parrinello and A. Rahman, *Physical Review Letters*, 1980, **45**, 1196.
48. M. P. Allen and D. J. Tildesley, *Computer simulation of liquids*, Oxford university press, 2017.
49. S. Nosé, *The Journal of chemical physics*, 1984, **81**, 511-519.
50. S. L. Chaplot, N. Choudhury, S. Ghose, R. Mittal and G. Prabhatasree, *European Journal of Mineralogy*, 2002, **14**, 291-329.
51. G. Cardini, P. Procacci and R. Righini, *Chemical physics*, 1987, **117**, 355-366.
52. J. Turney, E. Landry, A. McGaughey and C. Amon, *Physical Review B*, 2009, **79**, 064301.
53. A. Sartbaeva, S. A. Redfern and W. T. Lee, *Journal of Physics: Condensed Matter*, 2004, **16**, 5267.
54. B. Renker, H. Bernotat, G. Heger, N. Lehner and W. Press, *Solid State Ionics*, 1983, **9-10**, 1341-1343.
55. H. Mehrer, *Diffusion in solids: fundamentals, methods, materials, diffusion-controlled processes*, Springer Science & Business Media, 2007.
56. H. Böhm, *physica status solidi (a)*, 1975, **30**, 531-536.
57. A. Padilha, H. Raebiger, A. Rocha and G. Dalpian, *Scientific reports*, 2016, **6**, 28871.
58. O. Bierwagen, A. Proessdorf, M. Niehle, F. Grosse, A. Trampert and M. Klingsporn, *Crystal Growth & Design*, 2013, **13**, 3645-3650.
59. R. Chatten, A. V. Chadwick, A. Rougier and P. J. Lindan, *The Journal of Physical Chemistry B*, 2005, **109**, 3146-3156.
60. A. Lichtenstein, R. Jones, H. Xu and P. Heaney, *Physical Review B*, 1998, **58**, 6219.

TABLE-I. Comparison of enthalpy (H) per formula unit (f.u.) and lattice parameters (a & c) for some of the possible structural configurations with different distribution of Li atoms in the hexagonal channels of $\beta$-eucryptite. $N_A$ and $N_S$ represent the number of A and S type Li channels respectively in single unit cell containing 84 atoms. These structures are relaxed for all the degree of freedoms under the framework of DFT. $\Delta H$ represents the difference in enthalpy (meV/atom) with respect to that of the RT-phase.

| Sr. No. | Phase Description | $N_A$: $N_S$ | H(eV)/f.u. | $\Delta$H(meV)/f.u. | a(Å) | c(Å) |
|---|---|---|---|---|---|---|
| (i) | RT- Phase | 1:3 | -50.6072 | 0 | 10.5735 | 11.4100 |
| (ii) | All Li in Si Plane | 0:4 | -50.5981 | 9.1 | 10.5702 | 11.4763 |
| (iii) | Equal number of Li in Al and Si planes | 2:2 | -50.5953 | 11.9 | 10.5772 | 11.3156 |
| (iv) | One Li channel coplanar with Si | 3:1 | -50.5680 | 39.2 | 10.6317 | 11.2108 |
| (v) | All Li in Al plane | 4:0 | -50.5064 | 100.8 | 10.6934 | 11.1189 |
| (vi) | Li at 0, c/6 and c/3 in One channel. | | -50.3937 | 213.15 | 10.5252 | 11.5990 |
| (vii) | Li at 0, c/6 and c/3 in all channels. | | -49.5628 | 1044.4 | 10.4656 | 11.7154 |

FIG. 1: (Colour online) (a)The calculated (AIMD, this work) and experimental phonon spectra[35] of *β*-eucryptite at 300 K. (b)The calculated and experimental[60] (neutron diffraction) lattice parameters of *β*-eucryptite.

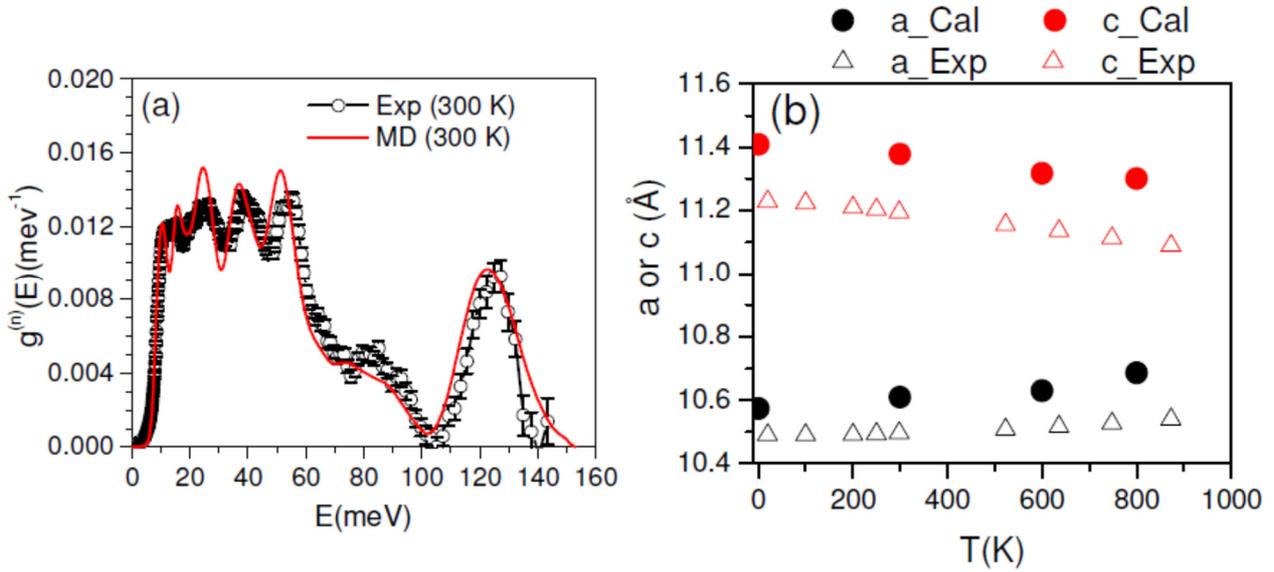

FIG. 2: (Colour online) The structure of RT phase of *β*-eucryptite (a) projected in a-b plane, and (b) containing three type of Li arranged in two types of channels, black Li forms A- type while pink and green Li form S- type channels. The structure of HT phase of *β*-eucryptite projected (a) in a-b plane, and (b) along c-axis. HT-Phase contains only one type of Li atoms. Key: $AlO_4$- Blue & $SiO_4$-Red.

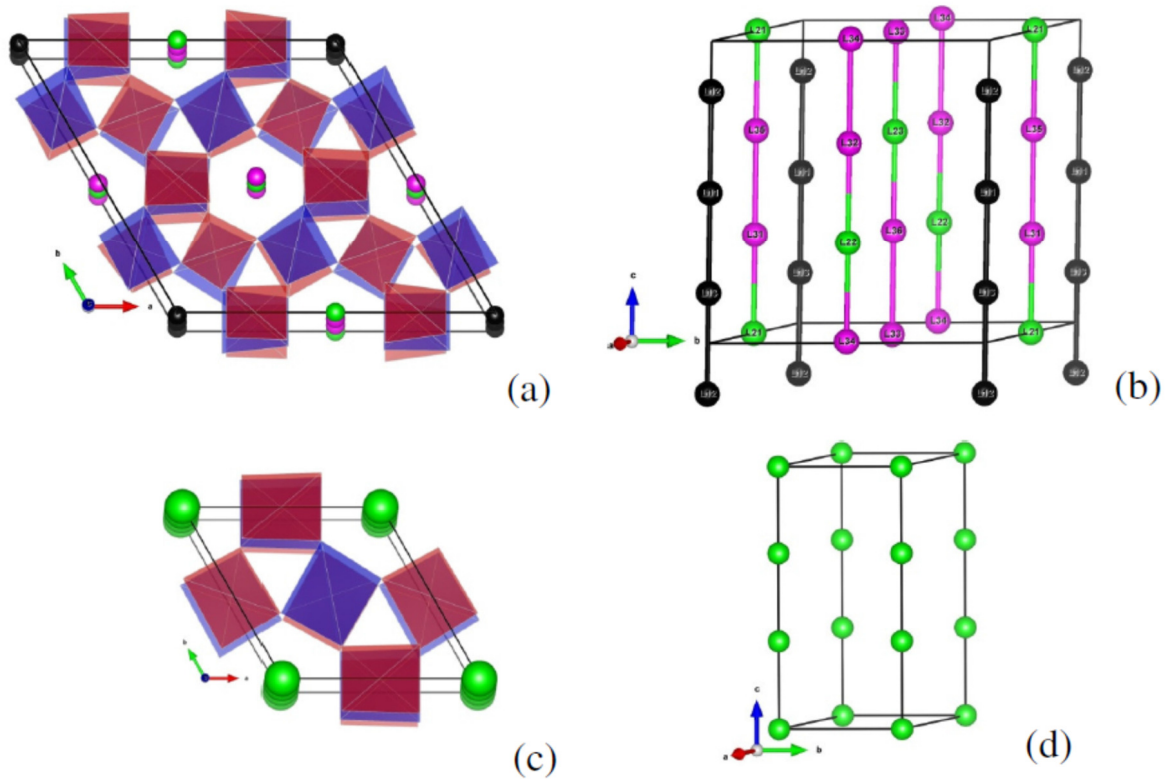

FIG. 3: (Colour online) The AIMD calculated mean square displacements of various atoms of RT-phase of β-eucryptite as a function of time (a) at 300K and (b) 600K. (c) Anisotropic MSD of Li averaged over all the Li atoms in the supercell and (d) MSD of each individual Li atoms along c-axis at 600K as a function of time.

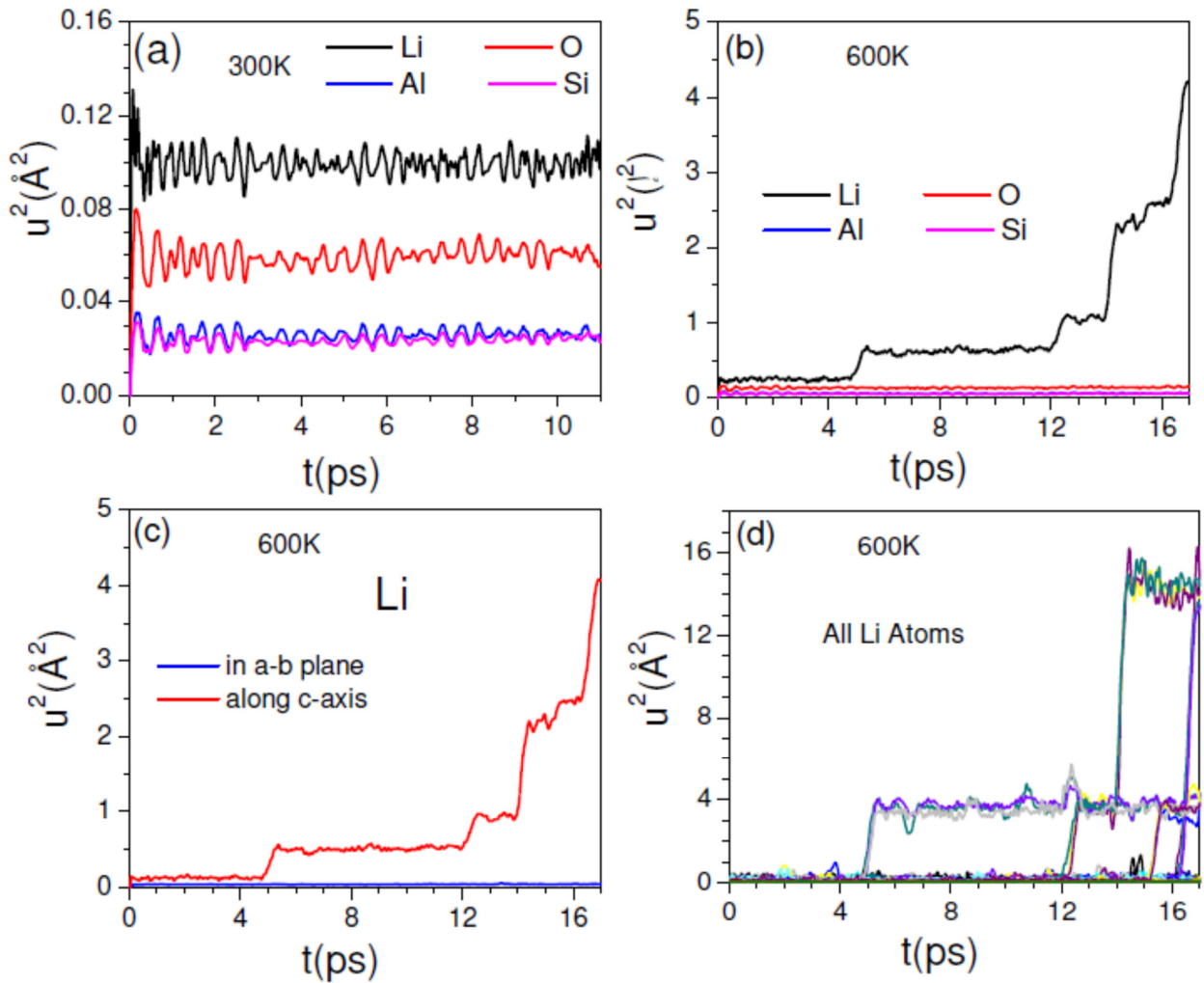

FIG. 4: (Colour Online) Trajectory of Li atoms in channels of β-eucryptite (RT Phase) at temperatures like (a) 300K, (b) 600K. Key: AlO$_4$-Blue, SiO$_4$-Red, Initial Position of Li- Black, Time Evolution of Li- Green balls.

(a) T=300K, RT Phase Unit Cell

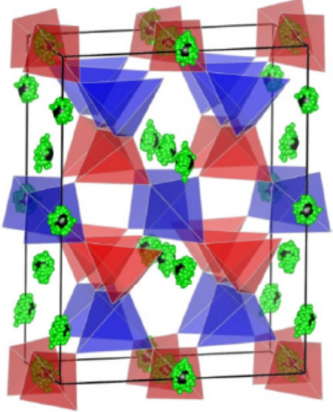
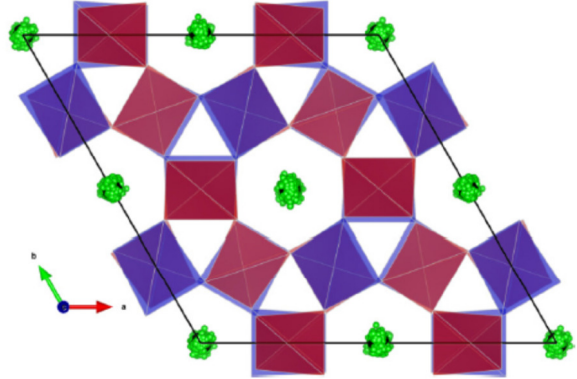

(b) T=600K, RT Phase Unit Cell

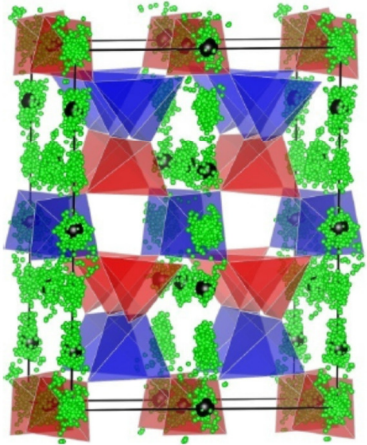
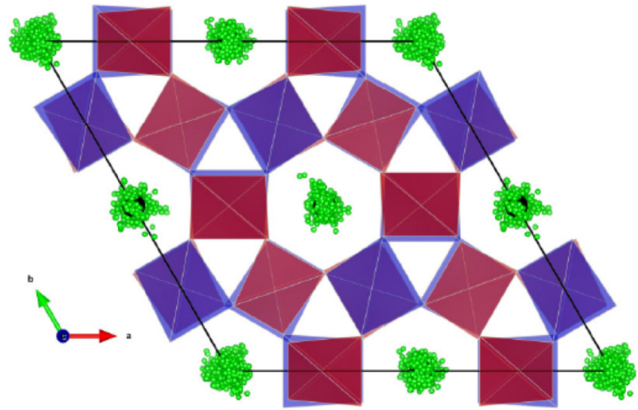

FIG. 5: (Colour online) The AIMD calculated mean square displacements of each individual Li atoms of HT- phase of $\beta$-eucryptite as a function of time along c-axis at (a) 800K (in 3×3×1 supercell), (b) 1000K (in 3×3×1 supercell) and (c) 1000K (in 2×2×2 supercell).

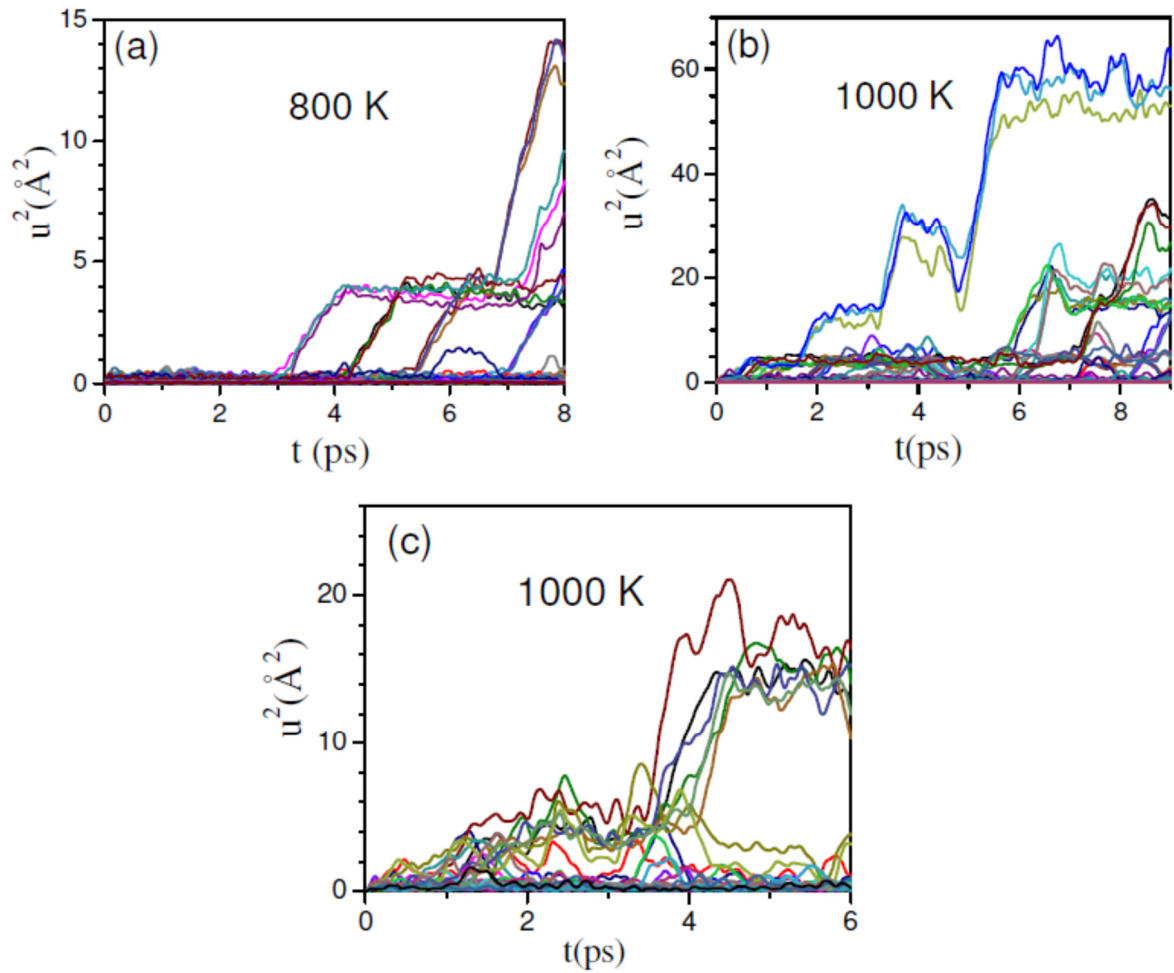

FIG. 6. (Colour Online): Trajectory of Li atoms in channels of *β*-eucryptite (HT-Phase) at various temperatures like (a) 800K, (b) 1000K and (c) 1200K. Key: AlO$_4$-Blue, SiO$_4$-Red, Initial Position of Li- Black, Time Evolution of Li- Green balls.

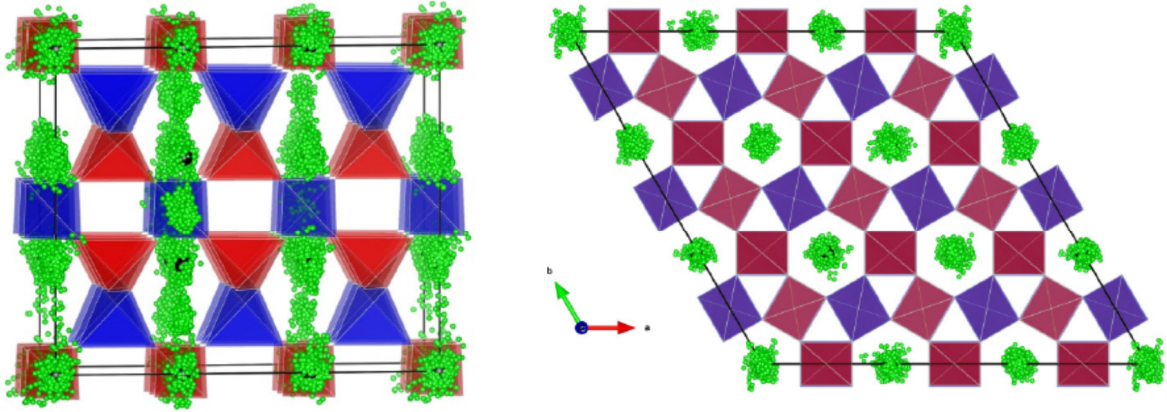

(a) T=800K, HT Phase Supercell (3×3×1)

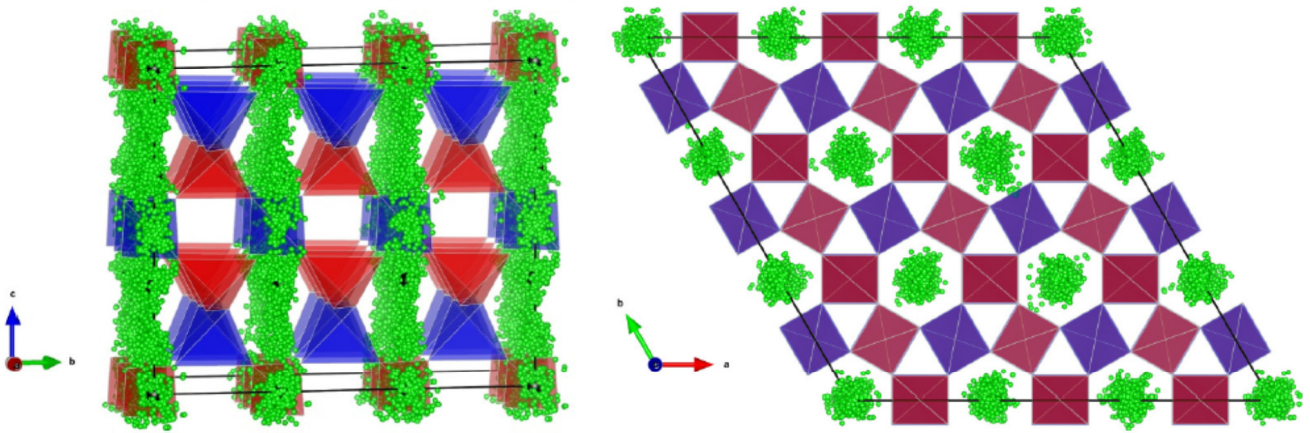

(b) T=1000K, HT Phase Supercell (3×3×1)

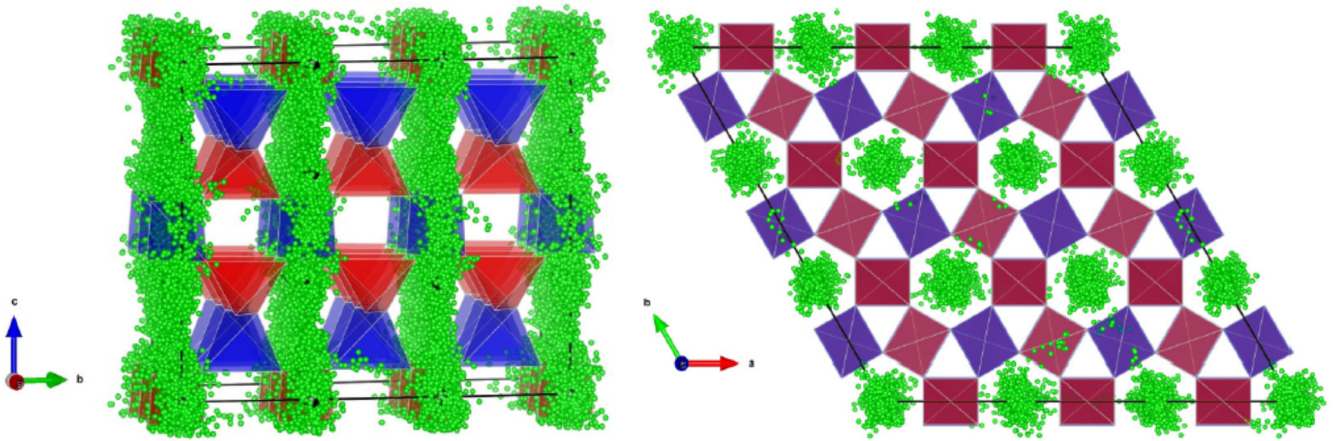

(c) T=1200K, HT Phase Supercell (3×3×1)

FIG. 7: (Colour online) The AIMD calculated (a) one dimensional diffusion coefficient as a function of temperature and (b) Arrhenius plot of diffusion coefficient and temperature to obtain activation energy for Li diffusion.

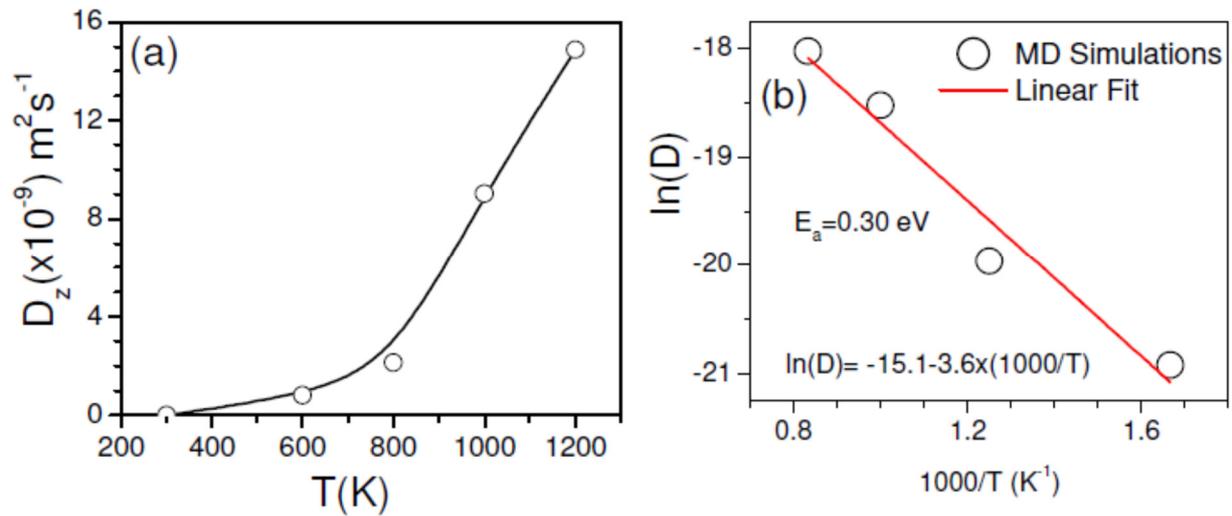

FIG. 8: (Colour online): The AIMD calculated temperature dependence of pair distribution function for various atomic pair of β-eucryptite. The information about different coloured lines is valid for all the subplots.

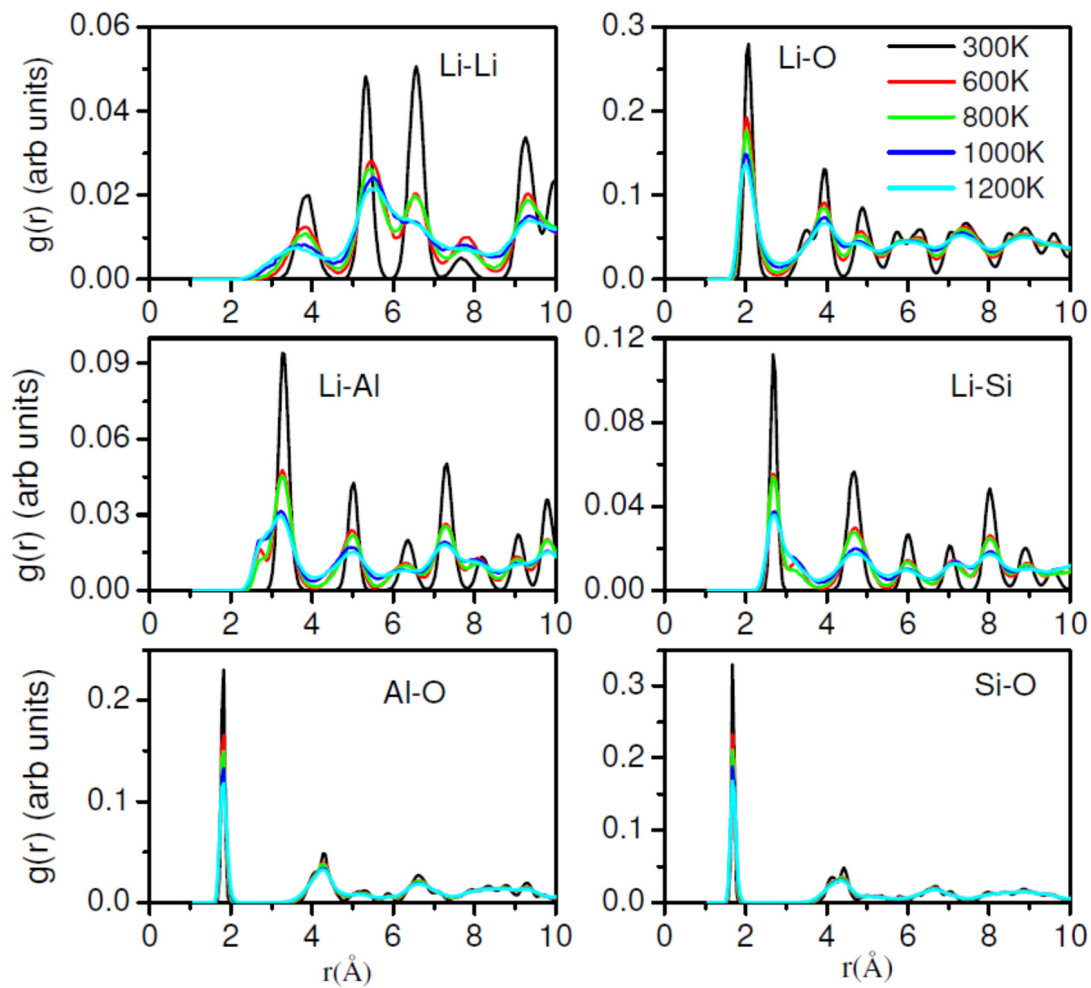

FIG. 9: (Colour online): The AIMD calculated temperature dependence of angular distribution function for various polyhedral angles in *β*-eucryptite. The information about different coloured lines is valid for all the subplots.

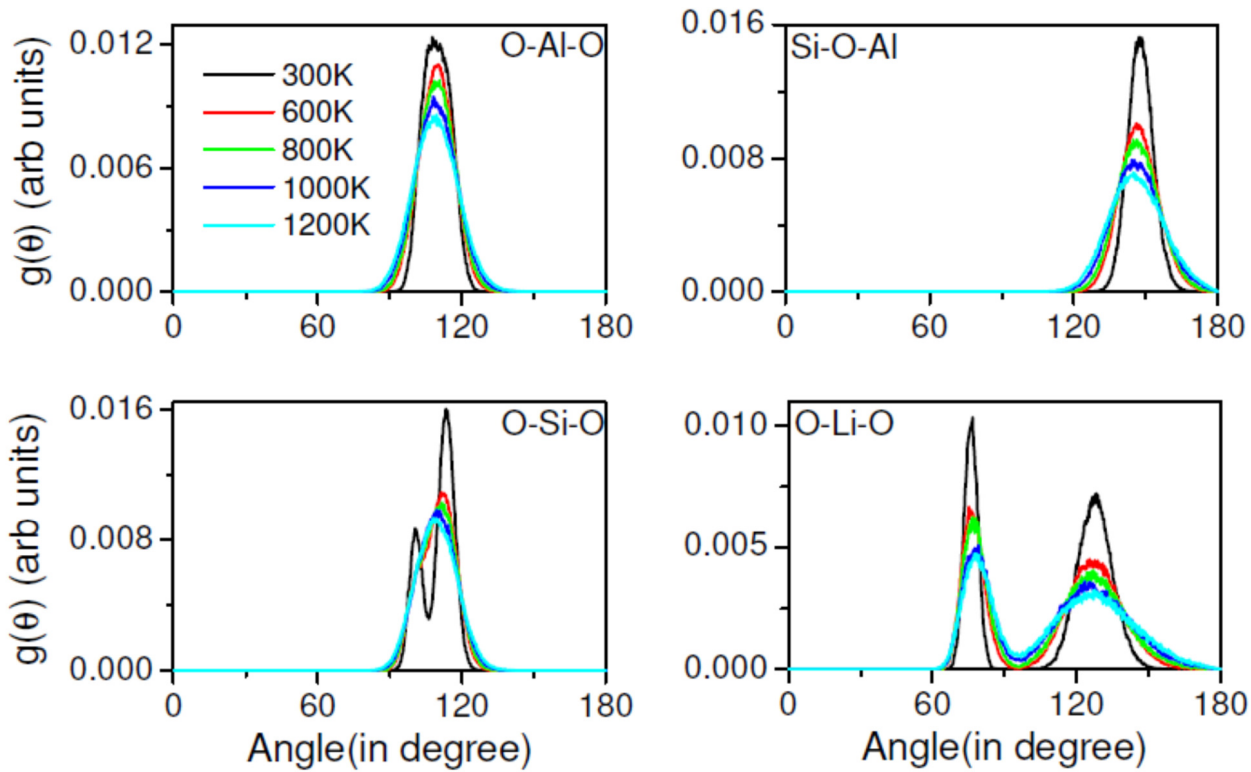

FIG. 10: (Colour online): (left) A cartoon showing the effect of decreasing c/a ratio on structure of *β*-eucryptite and (right) AIMD calculated time evolution of MSD's of Li atoms averaged over all the atoms in the super cell of HT- phase of *β*-eucryptite at various c/a ratio at T=800K.

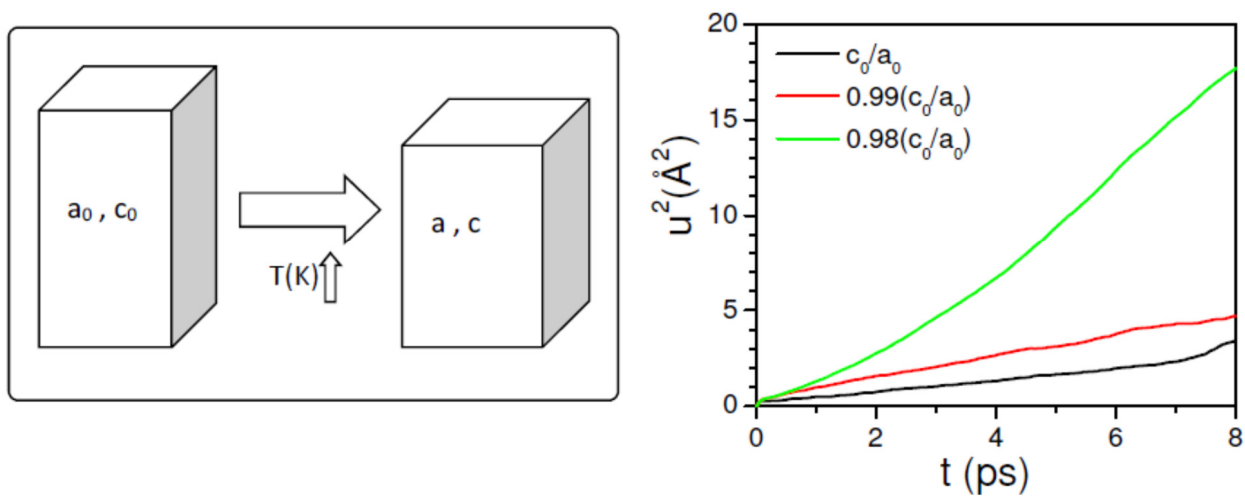

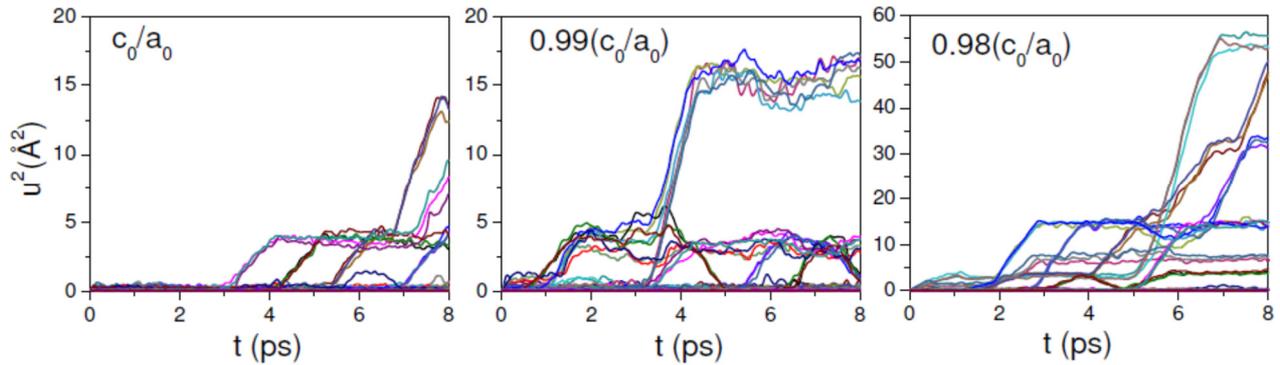

FIG.11: (Colour online): The AIMD calculated time evolution of MSD's of each individual Li along the hexagonal c-axis in the supercell (3×3×1) of HT- phase of β-eucryptite at various c/a ratios at T=800K.

FIG. 12: (Colour online): The ab initio DFT calculated static activation energy barrier for Li channel diffusion in HT phase of β-eucryptite at various c/a ratios. The calculations are performed by using climbing image nudged elastic band method.

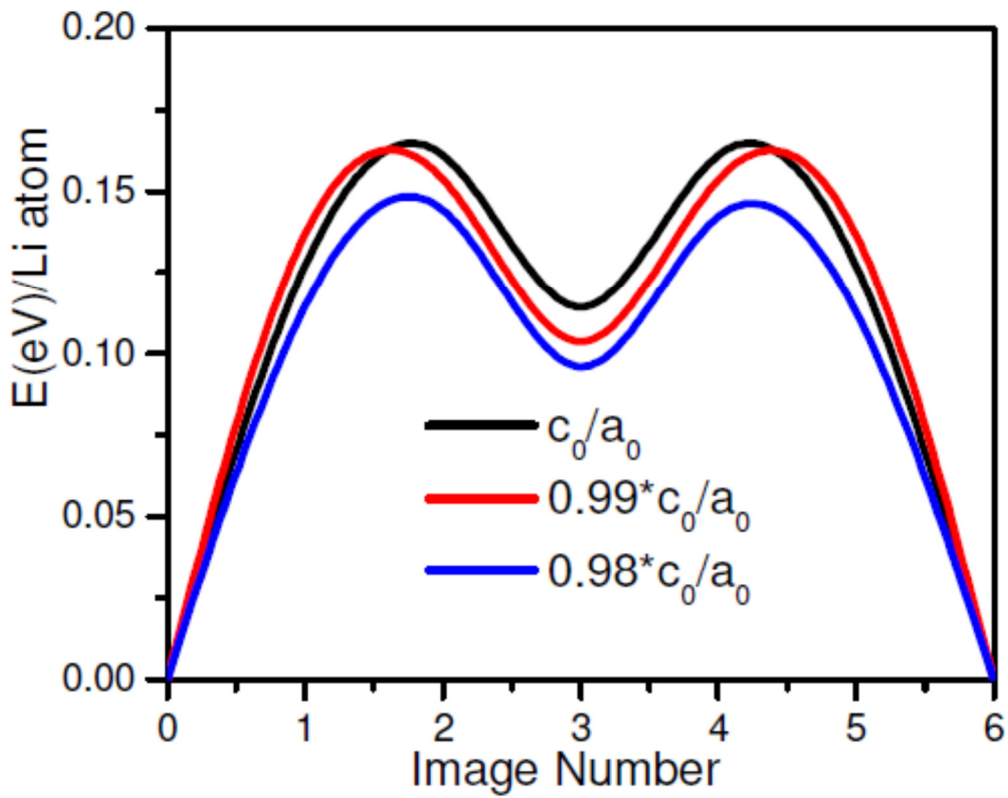

FIG. 13: (Colour online): The AIMD calculated time evolution of anisotropic MSD's of each individual Li in the supercell (3×3×1) of HT- phase containing one additional Li at tetrahedral void in the central channel at 800K.

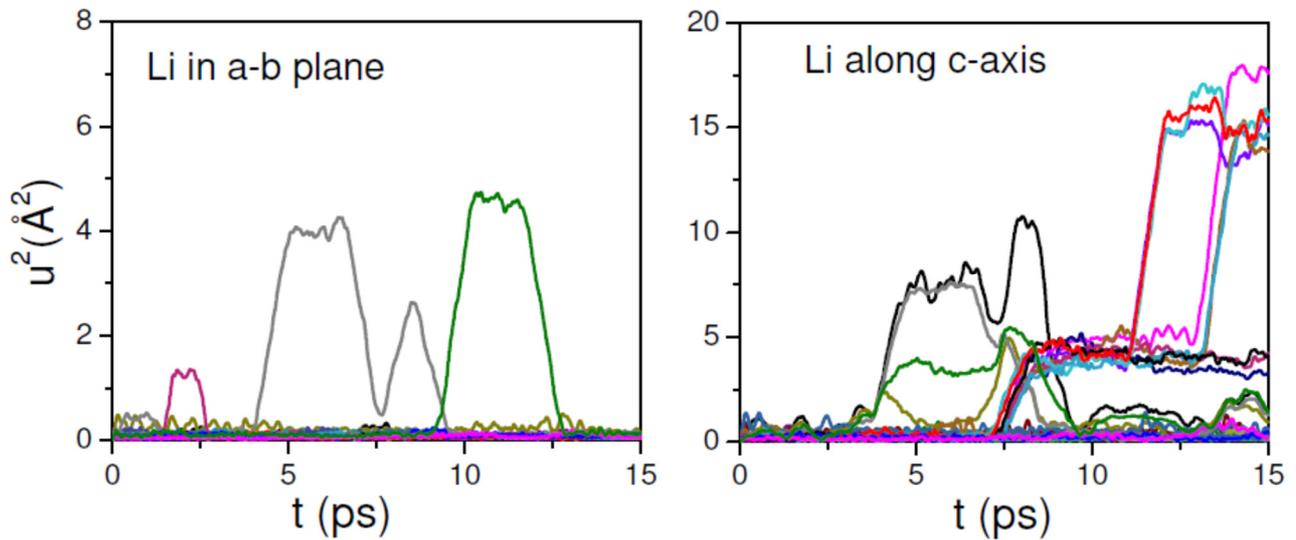

FIG. 14: (Colour online): The AIMD calculated time evolution of anisotropic MSD's of each individual Li in the super cell (3×3×1) of HT- phase containing one Li vacancy in the central channel at 800K.

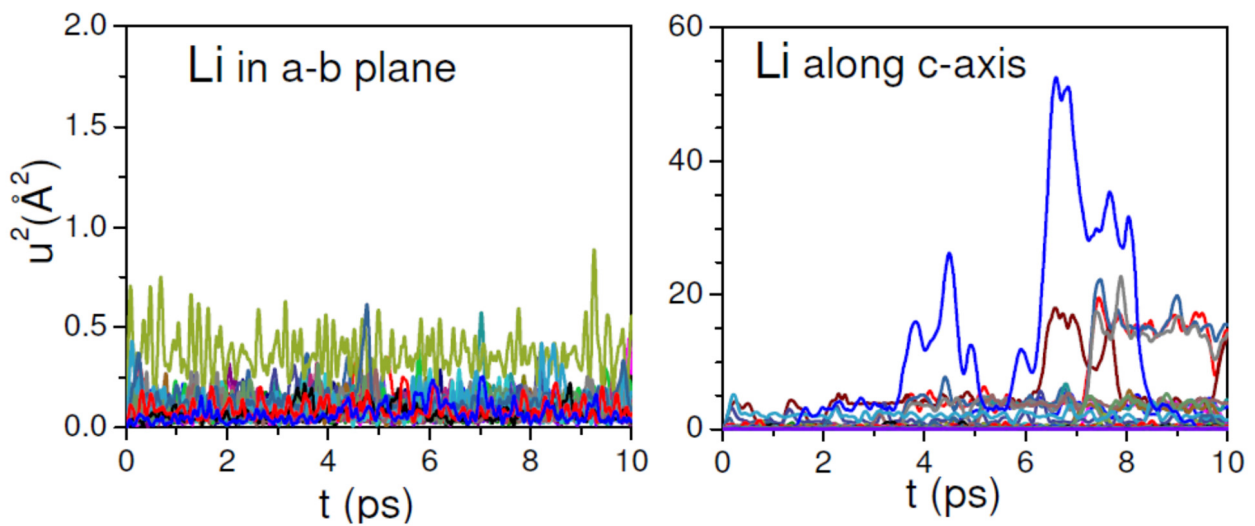

FIG.15: (Colour online): The AIMD calculated time evolution of anisotropic MSD's of each individual Li in the super cell (3×3×1) of HT- phase containing one O vacancy near the centre of the cell at 800K.

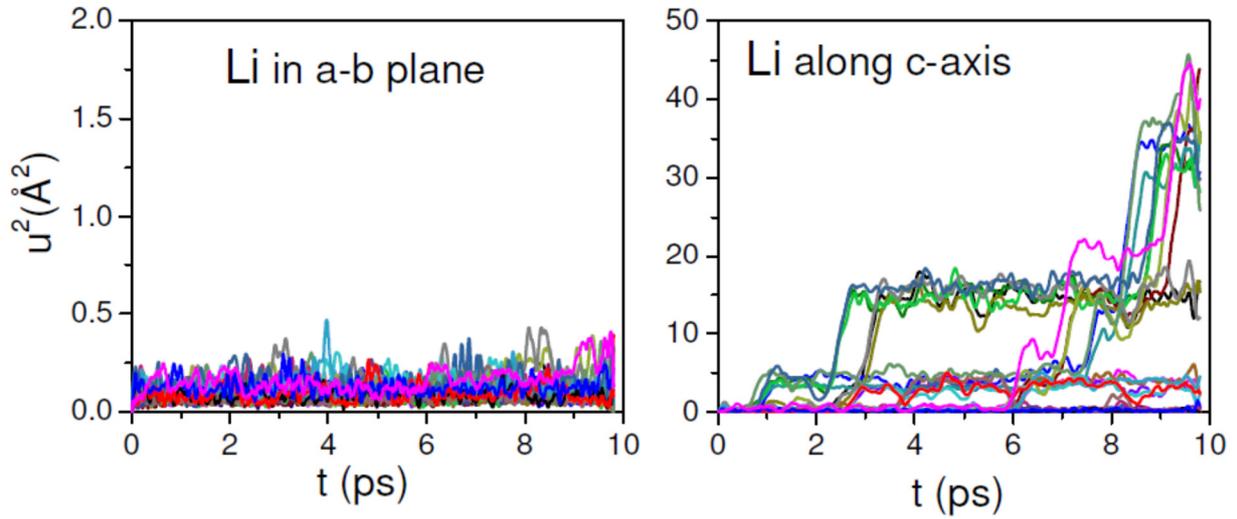